\documentclass[prd,10pt,aps,twocolumn,notitlepage,amsmath,nofootinbib,superscriptaddress,altaffilletter,showpacs,showkeys]{revtex4-2}

\usepackage{amsmath,amssymb,amsfonts,amstext}

\usepackage{graphicx}
\usepackage{rotating}
\usepackage{array}
\usepackage{booktabs}
\usepackage{float}

\usepackage[english]{babel}
\usepackage{textcomp}
\usepackage{indentfirst}
\usepackage{microtype}

\usepackage{bbm}
\usepackage{dsfont}
\usepackage{braket}

\usepackage[svgnames,table]{xcolor}
\usepackage[font=small]{caption}
\usepackage{url}
\usepackage{hyperref}
\hypersetup{
    colorlinks=true,
    linkcolor=red,
    citecolor=blue,
    urlcolor=red
}

\begin{document}

\title{Universality in quasinormal modes of a magnetized black hole}
\author{Marcos R. Ribeiro}\email{marcosribeiro@usp.br}
\affiliation{Department of Mathematical Physics, Institute of Physics, University of S\~ao Paulo,
05314-970 S\~ao Paulo, Brazil.}
\author{Eveling C. Ribeiro}\email{evelingmilena@usp.br}
\affiliation{Department of General Physics, Institute of Physics, University of S\~ao Paulo,
05314-970 S\~ao Paulo, Brazil}
\author{Kai Lin}\email{kailin@if.usp.br}
\affiliation{Federal University of Campina Grande, Campina Grande, PB 58429-900, Brasil}
\author{Elcio Abdalla}\email{eabdalla@usp.br}
\affiliation{Department of General Physics, Institute of Physics, University of S\~ao Paulo,
05314-970 S\~ao Paulo, Brazil}
\affiliation{Department of Physics, Center for Exact and Natural Sciences, Federal University of Para\'iba, 58059-970, Jo\~ao Pessoa, Brazil}
\affiliation{Para\'iba State University, 351 Bara\'unas Street, University District, Campina Grande, Brazil}

\bibliographystyle{plain}
 
\begin{abstract}
 In this work, we investigate the linear stability of a magnetized Einstein-Maxwell solution describing a static, axially symmetric black hole (BH) immersed in a uniform magnetic field $B$. We probe the dynamics of an external charged scalar field through its quasinormal modes (QNMs), combining frequency- and time-domain analyses. We find a critical value of the field charge at which the QNM spectrum exhibits universal power-law scaling with an exponent of approximately $1/2$. This critical behavior admits a simple interpretation in terms of a transition between a confined regime, where waves remain effectively trapped within a region of characteristic size $\sim 1/B$, and a deconfined regime, where the field reaches distances $\gg 1/B$ and the damping rate becomes parametrically small. These results provide qualitative and quantitative insights that may inform more realistic scenarios involving highly magnetized compact objects.

\end{abstract}
\date{\today}
\maketitle

\section{Introduction}
A deep understanding of gravity remains a central goal of theoretical physics. In the classical regime, spacetime dynamics is governed by general relativity via the Einstein field equations, which admit a wide variety of solutions of direct astrophysical relevance. For instance, the Schwarzschild solution and its generalizations describe spacetime around compact, spherically symmetric bodies with distinct properties \cite{schwarzschild1999gravitationalfieldmasspoint, stephani2009exact, griffiths2009exact}, enabling theoretical studies of astrophysical objects such as stars and BHs.

Remarkably, BH have shifted from a purely theoretical subject to an important 
focus of observational astrophysics, especially with the detection of gravitational
waves by the LIGO--Virgo Collaboration \cite{Abbott_2016}.
 Such signals are expected not only from neutron-star collisions and BH 
mergers but also from primordial processes in the early Universe \cite{gravitationalwavesfrominflation_2016}. 
On the other hand, the Event Horizon Telescope team has provided images of the shadow of supermassive black holes at galactic centers \cite{akiyama2022first}. Alongside these advances,
 radio observations have also proven to be a powerful means of probing the
 environment of compact objects, since cataclysmic events often produce
 detectable afterglows \cite{hallinan2017radio}. In this context, the BINGO radio telescope \cite{abdalla2022bingo,dos2024bingo}
will provide a valuable opportunity to explore such radio signatures, 
contributing to a deeper understanding of the astrophysical processes 
associated with compact objects.

BHs are often surrounded by accretion disks, which are made mostly of charged matter. The stability of this system is of high astrophysical interest, as it underlies many phenomena \cite{blandford1982hydromagnetic}. A particularly relevant case arises when a magnetic field permeates the compact object environment. These configurations have been discussed in connection with explosive events such as fast radio bursts \cite{zhang2020physical, Bochenek_2020}.
A complete theoretical solution of this problem is challenging and must self-consistently incorporate both gravity and relativistic plasma dynamics \cite{korobkin2011stability,RevModPhys.70.1}. Nevertheless, simplified models based on exact solutions of the Einstein--Maxwell equations provide a valuable starting point. This approach has been used by many authors and has yielded a number of interesting results \cite{konoplya2006stability, Brito_2014, Filho:2024ilq, Lin:2019fte, Zhu_2014, Ribeiro:2024jkm, de_Freitas_2017,aly2025more}.

F. J. Ernst developed in the 1970s a method for constructing exact non-asymptotically flat solutions of the Einstein--Maxwell equations from a previously known solution \cite{1976JMP....17...54E}. Over the years, these spacetimes have attracted considerable attention due to their potential astrophysical applications \cite{Brito_2014,shaymatov2022constraints,nayak1997gyroscopic,konoplya2008quasinormal,Horowitz_1997}. The simplest example describes a static black hole immersed in a uniform magnetic field $B$, namely the Ernst-Schwarzschild (ES) solution. In particular, \cite{Brito_2014} showed that this solution is stable under massless scalar perturbations. Moreover, these fields are confined to a region of order $\sim 1/B$ due to the presence of the magnetic field. Despite these important results, Ernst spacetimes remain relatively unexplored given their great potential for further applications.

In general relativity, critical behavior and universality can emerge in a striking way, signaling qualitative changes in a system's dynamics when a control parameter reaches a threshold value. A prominent example is the gravitational collapse studied by Choptuik and others \cite{choptuik1993universality, werneck2021nrpycritcol}, where near-critical evolutions display universal scaling governed by a critical exponent. Related ideas also appear in black hole perturbation theory, where near-horizon geometry can exhibit an emergent conformal symmetry that organizes universal scaling patterns \cite{gralla2018scaling}. In the same context, it is well established that charged fields can cause instabilities under a variety of circumstances \cite{Zhu_2014,Filho:2024ilq}. This aspect is particularly relevant in astrophysical settings, where charged fields may naturally occur in the vicinity of realistic BHs. The propagation of charged scalar fields in the ES background was analyzed in \cite{becar2023quasinormal}. Nevertheless, that study was restricted to the regime in which the dimensionless parameter $Br$ (with $r$ being the radial coordinate) remains very small, thus neglecting the non-trivial asymptotic features inherent to the ES spacetime. 

In this work, we investigate the propagation of charged scalar fields in the ES spacetime without relying on simplifying approximations. Our results show that, although magnetized BHs remain stable under charged scalar perturbations, the associated QNM spectrum exhibits a critical behavior at a threshold charge $q_c$. In the limit $q \to q_c$, the effective potential undergoes a qualitative topological change, leading to the deconfinement of the scalar field. We interpret this critical limit as a simplified mechanism that may serve as a toy model for emission processes.

The paper is organized as follows. In Section~\ref{II}, we briefly review the ES solution and its main features. In Section~\ref{III}, we study the dynamics of a charged scalar field in the ES spacetime and present the QNMs using two different approaches: direct integration (Sec.~\ref{IIIb}) and the time-domain profile (Sec.~\ref{IIIc}). Finally, in Section~\ref{IV}, we discuss our results and outline possible directions for future research.

\section{Black hole in a magnetic Field}\label{II}
The ES spacetime is an axisymmetric solution of the Einstein--Maxwell equations, first discussed in \cite{1976JMP....17...54E}. It describes a static BH immersed in a uniform magnetic field. In geometric units ($G=c=4\pi\epsilon_0=1$), the line element and the electromagnetic potential are given by
\begin{align}
\mathrm{d}s^2 &= \Lambda^2 \left( f(r)\mathrm{d}t^2 - \frac{\mathrm{d}r^2}{f(r)} - r^2 \mathrm{d}\theta^2 \right)
- \frac{r^2 \sin^2{\theta}}{\Lambda^2}\mathrm{d}\phi^2, \label{ernstmetric} \\
A_{\mu} &= - \frac{B r^2 \sin^2{\theta}}{\Lambda}\delta^3_{\mu}\label{ErnstA}
\end{align}
where $f(r)=1-{2M}/{r}$ and Greek indices running from $0$ to $3$ . The factor $\Lambda$ depends on the parameter $B$ as
\begin{equation}
\Lambda = 1 + B^2 r^2 \sin^2{\theta}. \label{lambda}
\end{equation}
Here, $B$ and $M$ denote the strength of the external magnetic field and the total mass of the BH, respectively. Accordingly, the natural dimensionless parameter characterizing the ES solution is $MB$.

It is easy to note that in the limit of no external field $B\to0$ we recover the Schwarzschild spacetime from the set of equations (\ref{ernstmetric}). In fact, the strength of the electromagnetic field acts as a deformation parameter of the spherical symmetry, and its inverse defines a characteristic length scale. This means that the gravitational effect of the electromagnetic field is weaker when $r \ll 1/B$, however, it increases with distance from the BH. When $r \gg 1/B$ the deformation of the Schwarzschild geometry is extremely strong and the solution (\ref{ernstmetric}) approximates the Melvin universe \cite{Melvin:1963qx, thorne1965absolute}. As a matter of fact, this feature ensures that the ES spacetime is not asymptotically flat when $B\ne0$. 

The external electromagnetic field does not change the position of the event horizon, which remains at $r=2M$, but makes the horizon no longer spherical due to the $\theta$-dependence of $\Lambda$ \cite{wild1980surface}. Indeed, in the axis of symmetry, the structure of the ES spacetime is similar to the Schwarzschild solution. On the other hand, out of the axis of symmetry we get a very peculiar geometry with many physical differences from the usual Schwarzschild BH \cite{dadhich1979trajectories, stuchlik1999photon}. For instance, in the equatorial plane, the Gaussian curvature of the event horizon is negative when $MB>1$ \cite{wild1980surface}.

\section{Charged scalar perturbation}\label{III}

\subsection{The perturbation equations}\label{IIIa}
We consider the dynamics, in the ES background, of a charged scalar perturbation field $\Psi$ which is minimally coupled to gravity via the action
\begin{equation}
    S_{\Psi}=-\frac{1}{2}\int\sqrt{-g}\mathrm{d}^4x\Big(g^{\alpha\beta}\nabla_{\alpha}\Psi^{*}\nabla_{\beta}\Psi+\mu^2\Psi^{*}\Psi\Big),
\end{equation}
where $g$ is the determinant of the ES metric $g_{\alpha\beta}$, and $\nabla_{\alpha}$ implements the minimal coupling between the charged field $\Psi$ and the ES electromagnetic potential (\ref{ErnstA}) via the replacement $\nabla_{\alpha} \to \nabla_{\alpha} + \mathrm{i}q A_{\alpha}$.

The variational principle leads us to the equation of motion for scalar fields, which is simply the Klein-Gordon equation,
\begin{equation}
    g^{\alpha \beta}\left(\nabla_{\alpha}+\mathrm{i}qA_{\alpha}\right)\left(\nabla_{\beta}+\mathrm{i}qA_{\beta}\right)\Psi + \mu ^2 \Psi = 0.\label{KGeq}
\end{equation}
\begin{figure*}[t]
\centering
\includegraphics[scale=0.77]{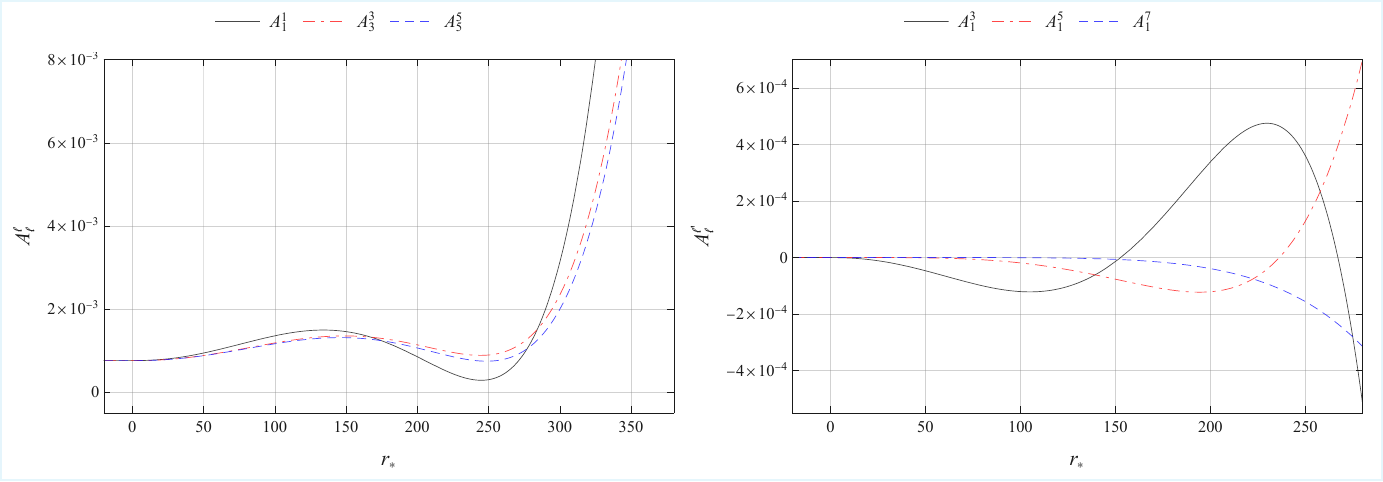}
\caption{Radial behavior of the matrix elements defined in Eq.~\eqref{Amatrix}, for \(MB=0.02\), \(q/q_c=1.05\), \(m=1\), and \(\mu=0\). The left panel shows the first diagonal elements, while the right panel shows the first off-diagonal ones.}
\label{matrix}
\end{figure*}
Even though spherical symmetry is broken, a decomposition in terms of spherical harmonics is still useful. Thus, let us consider the following ansatz:
\begin{equation}
    \Psi\left(t, r, \theta, \varphi\right) = \frac{1}{r}\sum_{\ell,m} \psi_{\ell m}\left(r,t\right)Y_{\ell}^m\left(\theta, \varphi \right). \label{SV}
\end{equation}
where $\ell$ is the multipole number ($\ell = 0, 1, 2, \ldots$) and $m$ is the azimuthal number with $|m| \leq \ell$. Moreover, consider the orthogonality condition for the spherical harmonics:
\begin{align}
     \langle Y^m_\ell,Y_{\ell'}^{m'} \rangle = \int \mathrm{d}\Omega\,  Y_\ell^m\left(\theta, \phi \right)Y_{\ell'}^{m'*}\left(\theta, \phi \right)= \delta_{\ell\ell'}\delta_{mm'},\label{eqorto}
\end{align}
where the integration is performed over the solid angle $\mathrm{d}\Omega$. Now, we can plug the ansatz (\ref{SV}) into Eq.~(\ref{KGeq}) to obtain the following system of equations for each mode $(\ell, m)$:

\begin{widetext}
\begin{align}
\frac{\partial^2 \psi_{\ell m}}{\partial r^2}
+ \frac{f'}{f} \frac{\partial \psi_{\ell m}}{\partial r}
- \frac{1}{f^2} \frac{\partial^2 \psi_{\ell m}}{\partial t^2}
- \frac{1}{f} \left(
    \frac{f'}{r}
    + \frac{\ell(\ell+1)}{r^2}
\right) \psi_{\ell m}= \frac{1}{f} \sum_{\ell'}A^{\ell'}_{\ell}\psi_{\ell'm}\label{mastereq}
\end{align}
\end{widetext}

The equation above is our master equation, which describes the dynamics of a charged scalar field in the ES spacetime. Note that our decomposition~(\ref{SV}) results in a system of differential equations, where $A_\ell^{\ell'}(r)$ couples different modes,
\begin{equation}
      A_\ell^{\ell'}(r):=f\langle Y_{\ell}^m,\Delta\left(r,\theta\right) Y_{\ell'}^m \rangle.\label{Amatrix}
\end{equation}
This is expected, as the magnetic field breaks the spherical symmetry. The explicit form of the function $\Delta(r,\theta)$ is given by:

\begin{align}
\Delta=\frac{m^2\left(\Lambda^4-1\right)}{r^2 \sin^2{\theta}} + \left(q^2B^2r^2 \sin^2{\theta}  + \mu^2 \right)\Lambda^2 -
2qmB\Lambda^3. \label{OMEGA}
\end{align}
The structure of the matrix $A_{\ell}^{\ell'}$, shown in Fig.~\ref{matrix}, ensures direct coupling only among the three nearest multipoles with the same parity. In other words, for fixed $\ell$, the six neighboring eigenfunctions $\{\psi_{\ell\pm2,m}, \psi_{\ell\pm4,m}, \psi_{\ell\pm6,m}\}$ belong to the same coupled sector. Although each mode couples directly only to its three nearest multipoles of the same parity, all modes within a given parity sector are indirectly coupled through successive interactions.

\begin{figure}[t]
\centering
\includegraphics[scale=0.56]{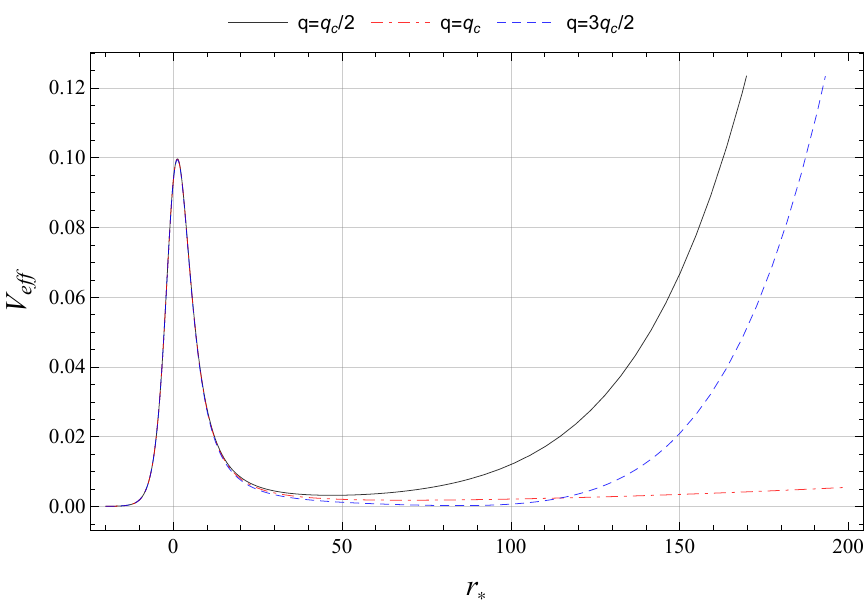}
\caption{Effective potential for different values of the perturbation charge $q$. We set $BM=0.02$, $\ell=m=1$, and $\mu=0$. Note that the growth at the critical charge $q_c$ (red dashed) is weaker.}
\label{veff_fig}
\end{figure}

One can perform a change of variable to tortoise coordinates $r_*=\int \mathrm{d}r/f(r)$ to obtain the wave-like equation from (\ref{mastereq}) as
\begin{equation}
    \frac{\partial^{2} \psi_{\ell m}}{\partial r^2_*}-
    \frac{\partial^{2} \psi_{\ell m}}{\partial t^2}= V_{\mathrm{eff}}\psi_{\ell m}+\sum_{\ell' \ne \ell}A^{\ell'}_{\ell}\psi_{\ell' m} \label{eqwave}
\end{equation}
where the effective potential is given by
\begin{equation}
    V_{\mathrm{eff}}= f\left(\frac{f'}{r}+\frac{\ell(\ell+1)}{r^2}\right)+A^{\ell}_{\ell}(r)\label{veff}
\end{equation}
It is worth noting that the potential above is symmetric under the combined transformation $m \to -m$ and $q \to -q$.
The effective potential has the standard barrier near the BH and is only weakly affected by the external magnetic field. However, the magnetic field becomes dominant in the asymptotic regime, as shown in Fig.~\ref{veff_fig}. In fact, the asymptotic growth
\begin{equation}  \lim_{r\to\infty}V_{\mathrm{eff}}\sim r^{6},
\end{equation}
ensures the confinement of any perturbation that satisfies (\ref{mastereq}). Nevertheless, due to the presence of electric charge in the scalar field, the asymptotic behavior can be smoothed out whenever $\mu=0$ and $q=q_c:=mB$. In this limit, the effective potential exhibits a quadratic growth analogous to the AdS case, with an effective AdS radius proportional to $1/B$. Moreover, the mode coupling becomes more localized in multipole space, in the sense that each mode couples only to its nearest neighbors of the same parity.
\begin{figure*}[htb]
\centering
\includegraphics[scale=0.92]{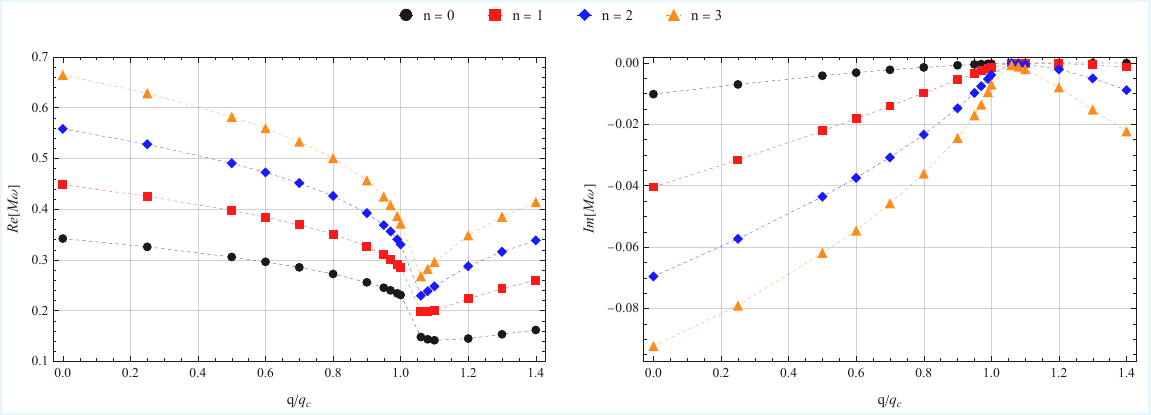}
\caption{Quasinormal spectrum of a massless perturbation as a function of the field charge normalized by the critical charge, $q_c$, for $MB=0.1$ and $\ell=m=1$. Two distinct regimes are observed, separated by the effective critical charge $Q_c$. When $q/q_c \in I_c$, the spectrum becomes more densely populated, making it harder to determine which modes belong to a given overtone sequence.}
\label{QNMresult}
\end{figure*}
\subsection{Frequency domain analysis}\label{IIIb}
In this section, we compute the QNM associated with the perturbation equation (\ref{mastereq}). 
It is worth noting that the presence of an effective infinite wall (Fig.~\ref{veff_fig}) 
leads to the appearance of long-lived modes in the quasinormal spectrum. 
To find these modes, we assume a simplified, decoupled case, namely 
$A^{\ell'}_{\ell} = 0$ for $\ell \neq \ell'$ in Eq.~(\ref{mastereq}). 
Our analysis is carried out in the frequency domain, adopting a separable ansatz of the form 
$\psi_{\ell m} = R_{\ell m}e^{-i\omega t}$. 
In this analysis, we also assume the following asymptotic behavior
\begin{equation}
R_{\ell m}(r) \approx 
\begin{cases}
e^{-\mathrm{i}\omega r_{*}} \quad\text{as}& r\to 2M \\
0, \quad \quad \ \ \text{as}& r\gg1/B
\end{cases} \label{BC}
\end{equation}
which represents an in-going wave at the event horizon and no waves far enough from BH.

The wave equation is given by
\begin{equation}
    f \frac{\mathrm{d}^2 R}{\mathrm{d} r^2}
    + f' \frac{\mathrm{d} R}{\mathrm{d} r}
    - \left(
        \frac{f'}{r}
        - \frac{\omega^2}{f}
        + \frac{\ell(\ell+1)}{r^2}
        + A^{\ell}_{\ell}
      \right) R = 0.
    \label{direct-eq}
\end{equation}
Our strategy is to apply a direct numerical integration of Eq.~(\ref{direct-eq}) over the interval
$I = [r_0, r_{\infty}]$, where the numerical parameters $r_0$ and $r_{\infty} \gg 1/B$ 
denote a region close to the black hole and a region far from it, respectively.

Close to the horizon, the solution behaves as
\begin{equation}
    R \sim \left(r-2M\right)^{-2\mathrm{i}M\omega}
    \sum_{n} a_{n} \left(r-2M\right)^{n},\label{nearHorizon}
\end{equation}
where $n > 0$, and the coefficients $a_n$ are determined in terms of $a_0 = 1$ by substituting the near-horizon expansion ~(\ref{nearHorizon}) into
Eq.~(\ref{direct-eq}) and solving order by order. The QNMs are obtained by imposing the second boundary condition given in Eq.~(\ref{BC}), which leads to an equation satisfied only by a discrete set of complex frequencies.

We performed a detailed numerical analysis of the QNM spectrum, which can be defined as a set of the kind $\omega_{n\ell m}$, as a function of $B$, and $q$, with representative results shown in Fig.~\ref{QNMresult}.  Our numerical scheme becomes less robust in the interval \(q/q_c \in I_c \approx [1,\,1.1]\), where an increase in the density of the mode makes the identification of individual branches more difficult. We observe a narrowing of this window on the right as the overtone index increases. For this reason, there is a small gap in Fig.~\ref{QNMresult}. In addition, we identify a special charge value $Q_c$, with $Q_c/q_c \in I_c$, at which the QNM behavior abruptly reverses. The real part of the quasinormal frequency (left panel of Fig.~\ref{QNMresult}) decreases with the field charge for $q < Q_c$ across all parameter values considered, whereas in the second regime, $q > Q_c$, it increases. Notably, both regimes obey a power-law scaling for the real frequency of the form

\begin{equation}
\mathrm{Re}[\omega] = \omega_c^{\pm}+\beta^{\pm} |q-Q_c|^{\gamma^{\pm}},\label{realfit}
\end{equation}
where $\omega_c$, $\beta$, $\gamma$, and $Q_c$ are positive numerical coefficients. We fit the two sides of the critical point separately. Here, the superscripts $\pm$ refer to the regimes $q<Q_c$ and $q>Q_c$, respectively, and $\omega_c^-$ and $\omega_c^+$ denote the corresponding critical frequencies from the left and right-sides. To avoid making premature assumptions, we do not impose that these quantities or the other fit coefficients be equal \emph{a priori}.

\begin{table}[t]
\renewcommand{\arraystretch}{1.1}    
\setlength{\tabcolsep}{8pt}          
\centering
\begin{tabular}{lccc}
  \hline\hline
  $n$ & $\gamma^{-}$ & $\gamma^{+}$ & $Q_c$ \\
  \hline
   0 & $0.50$ &        & $0.1046$ \\
   1 & $0.42$ &        & $0.1028$ \\
   2 & $0.42$ & $0.52$ & $0.1020$ \\
   3 & $0.44$ & $0.45$ & $0.1016$ \\
  \hline\hline
\end{tabular}
\caption{Numerical values for the effective critical charge and exponents for $MB=0.1$. In this analysis, we obtained errors of at most $7\%$ and $0.5\%$ for $\gamma^{\pm}$ and $Q_c$ respectively. Our numerical analysis is less stable on the right side, especially for small $n$, which explains the absence of these values.
}\label{Tabexpoentes}
\end{table}
We observe that $Q_c$ is not a single value, rather, it depends on the mode. We find a strong dependence on both the azimuthal number and the magnetic-field parameter. This behavior is expected since $Q_c$ represents a small deviation from the critical charge $q_c=mB$. For instance, $Q_c$ is approximately $2.4$ times larger when comparing the modes $\omega_{021}$ and $\omega_{022}$, with $B$ fixed. On the other hand, we find little dependence on $\ell$ or $n$. Remarkably, $Q_c$ approaches $q_c$ as the overtone number $n$ increases, as shown in Table~\ref{Tabexpoentes}.

The numerical calculations for the exponents $\gamma^{\pm}$ are also reported in Table~\ref{Tabexpoentes} for the case $MB=0.1$. However, for all modes tested at different values of $MB \ll 1$, the exponent remains within the error bars across these cases. This numerical evidence motivates us to regard it as universal, and we obtain a remarkably good fit in both regimes ($q\gtrless Q_c$), with $\gamma^{\pm}\approx 0.46$.
\begin{figure*}[htb]
\centering
\includegraphics[scale=0.92]{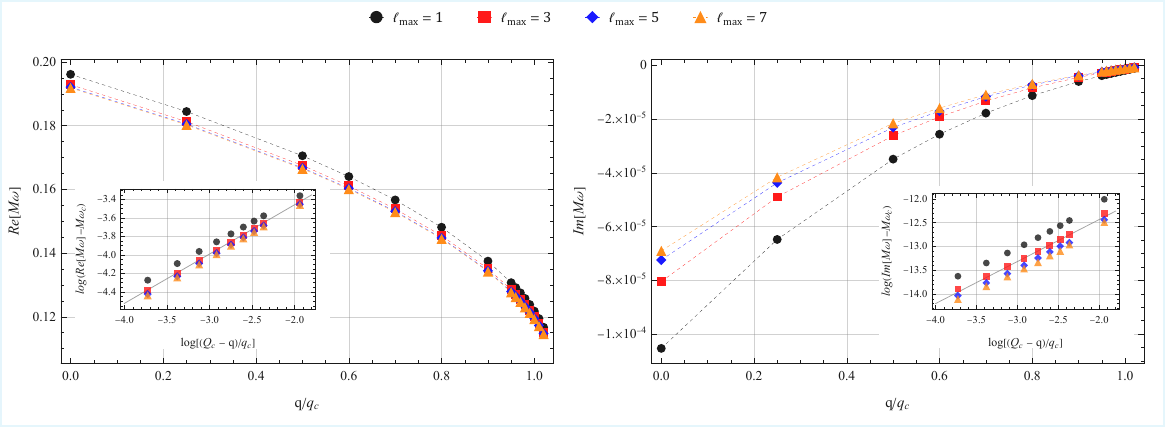}
\caption{Coupled QNM of the mode with $\ell=m=1$, namely $\Omega^{1}_{01}$, for $MB=0.05$ and different values of the multipole cutoff number. The same general scaling behavior, described by Eqs.~\eqref{realfit} and \eqref{imaginaryfit}, is still observed. The frequency follows the specific behavior near the critical point $q\approx Q_c$: $M\Omega^{1}_{01}\approx 0.1 + 0.09\, \mathfrak{q}^{0.54}-2\times 10^{-5}\mathfrak{q}^{0.88}\,\mathrm{i}$, where $\mathfrak{q}:=(Q_c-q)/q_c$. The insets show the linear behavior in the log-log plots, with $\gamma^{-} \approx 0.54$ for the real part (left) and $\xi^{-}\approx0.88$ for the imaginary part (right). We can see that the impact of the cutoff is stronger on the imaginary part.}
\label{n=0_B=0.05_lmax}
\end{figure*}

We observe that, at the critical point, the spectrum seems to bifurcate and new modes emerge. As a result, there is a natural gap in our system, which in turn requires distinct values for the numerical parameters for the different regimes. Specially, the critical frequency in the first regime is consistent with the following behavior
\begin{equation}
    \omega^{-}_c\sim\left(0.35n+0.036\ell+1.6m+0.39\right)B. \label{criticalomega}
\end{equation}
On the second-regime side, our numerical calculation is less stable and the associated uncertainties are significantly larger than those in the first regime ($q<Q_c$), so the data do not provide enough sensitivity to reliably determine the parameter dependence of $\omega_c^+$ in this regime. Despite this, within the parameter space explored, the critical frequency does not show significant variation and remains approximately constant. For $MB=0.1$ we found it around $\omega_c^{+}\approx 0.19$. The parameter $\beta$ is always of order of unity, increasing with $B$ and $n$ and decreasing with $m$.
\begin{table}[t]
\renewcommand{\arraystretch}{1.1}    
\setlength{\tabcolsep}{8pt}          
\centering
\begin{tabular}{lccc}
  \hline\hline
  $n$ & $\xi^{-}$ & $\xi^{+}$ & $Q_c$ \\
  \hline
   0 & $0.96$ &        & $0.1046$ \\
   1 & $0.90$ &        & $0.1028$ \\
   2 & $0.74$ & $1.89$ & $0.1020$ \\
   3 & $0.62$ & $1.78$ & $0.1016$ \\
  \hline\hline
\end{tabular}
\caption{Numerical values for the effective critical charge and exponents for $MB=0.1$ with $\ell=m=1$. In this analysis, we obtained errors of at most $4\%$. The reason for omitting the initial $n$ values is the same as in the previous table.
}\label{Tabexpoenteim}
\end{table}

The behavior of the imaginary part, shown on the right in Fig.~\ref{QNMresult}, exhibits a similar change in behavior in the transition from the first to the second regime. In particular, very close to $Q_c$, all modes become approximately equal and display a very small decay rate, indicating a quasi-degenerate state of the system. The imaginary part also follows a power-law behavior of the form
\begin{equation}
\mathrm{Im}[\omega]=-\zeta^{\pm}\left|q-Q_c\right|^{\xi^{\pm}},\label{imaginaryfit}
\end{equation}
where $\zeta$ and $\xi$ are positive numerical parameters. Here, $Q_c$ denotes the same effective critical charge as in the real part. Unlike $\mathrm{Re}[\omega]$, the exponent that governs the imaginary contribution depends non-trivially on $n$, $\ell$, and $m$. For the special case $\ell=m=1$, Fig.~\ref{QNMresult} shows a clear change in the concavity of the mode curve for different overtone numbers in the first regime. Indeed, we observe a small decrease in exponent $\xi^{-}$, with a possible large plateau $n$, similar to what we found for $Q_c$. That is, we expect the exponent to satisfy a relation of the form
\begin{equation}
\xi^{\pm}=\xi^{\pm}_{0}+ f(n,\ell,m),
\end{equation}
such that $\lim_{n\gg 1} f = 0$. However, due to the lack of numerical results in $n\gg1$, we cannot reliably determine the asymptotic value or the functional form of $f$. Assuming a rational dependence, we obtain for the first regime, from the data in Fig.~\ref{QNMresult},
\begin{equation}
\xi^{-}\approx0.34+\frac{1.1}{1+n}.
\end{equation}
\subsubsection*{Effect of mode coupling}
Here, we present the QNMs of the coupled system described by Eq.~(\ref{mastereq}). We introduce a multipole cutoff \(\ell_{\max}\), such that $A_{\ell}^{\ell'}=0$ for $\ell' > \ell_{\max}$. The resulting coupled system can then be treated within the more sophisticated framework described in \cite{pani2013advanced}. Our results were obtained using a generalized version of the direct integration method employed in the decoupled case.

At first sight, in the coupled problem, the multipole number $\ell$ is no longer a good label for the modes. Nevertheless, our analysis indicates that, at least in the regime $MB\ll 1$, $\ell$ still retains a clear physical meaning. In particular, we find that the decoupled families of modes with fixed multipole number are only slightly deformed by the coupling, as illustrated in Fig.~\ref{n=0_B=0.05_lmax}. Therefore, although all multipoles with the same parity belong to a common coupled spectral problem, the corresponding branches remain connected to the decoupled $\ell$-families. Symbolically, we write
\begin{equation}
   \Omega^{\ell}_{nm} = \omega_{n\ell m} + \delta\omega_{n\ell m},
\end{equation}
where $\Omega^{\ell}_{nm}$ denotes the coupled QNM connected to the decoupled mode $\omega_{n\ell m}$, and $\delta\omega_{n\ell m}$ represents the shift induced by mode coupling. In this sense, although multipoles with the same parity belong to a common coupled quasinormal spectrum, each branch can still be meaningfully classified by the labels $(n,\ell,m)$.

With this interpretation, we find that mode coupling produces predominantly quantitative corrections to the QNM spectrum, while preserving the same power-law scaling with the charge parameter, as illustrated in Fig.~\ref{n=0_B=0.05_lmax}. As expected, the approximation becomes more accurate as higher multipoles are included. Nevertheless, our results indicate that $\delta\omega$ already converges rapidly for relatively small values of $\ell_{\max}$. In particular, for the fundamental mode associated with $\ell=1$, increasing the cutoff from $\ell_{\max}=5$ to $\ell_{\max}=7$ modifies the real part by only about $0.1\%$, while the imaginary part changes by about $5\%$. Moreover, at the critical charge $q=q_c$, the coupling matrix $A_{\ell}^{\ell'}$ becomes effectively localized in $\ell$-space, with the dominant off-diagonal terms mainly restricted to the nearest neighbors, $\ell\pm2$. This explains the rapid convergence for relatively small $\ell_{\max}$ and shows that the decoupled approximation captures the leading critical behavior of the real part of the frequency, although nearest-neighbor coupling may still affect damping rates and late-time dynamics.

The numerical parameters in Eq.~\eqref{realfit} are only weakly affected by coupling. The largest variation is found for $\beta^{-}$, at the level of about $7\%$. Moreover, a single critical exponent, $\gamma^{-}\approx 0.54$, provides an excellent description of all datasets shown in Fig.~\ref{n=0_B=0.05_lmax}. This value is also consistent with that obtained in the decoupled case for $MB=0.1$, since the relative difference remains within our numerical uncertainty.

The power-law scaling associated with the damping rate, Eq.~(\ref{imaginaryfit}), exhibits a stronger dependence on its numerical parameters. For instance, the parameter $\zeta^{-}$ changes by about $40\%$ from the decoupled case to the ($\ell_{\max}=7$)--coupled case. On the other hand, the exponent \(\xi^{-}\)  shows little dependence on the multipole cutoff. However, in comparison with the \(MB=0.1\) case presented in Table~\ref{Tabexpoenteim}, we find a difference of about \(9\%\), suggesting a potentially more intricate dependence on the magnetic field than that observed for the real part.

Finally, it is important to emphasize that, although the numerical fits are obtained using only the subset of data closest to the effective critical point $Q_c$, the scaling law for the real part accurately reproduces the entire dataset in the region $0 \le q < Q_c$. By contrast, the scaling law for the imaginary part describes the behavior of the damping rates only in the vicinity of $Q_c$.
\subsection{Time domain analysis}
\label{IIIc}
\begin{figure*}[htb]
\centering
\includegraphics[scale=1.0]{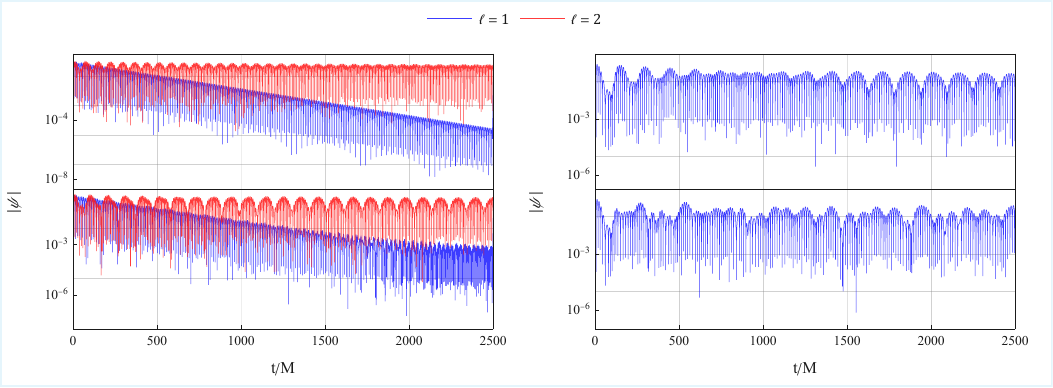}
\caption{
Time-domain profiles for $m=1$ and $MB=0.1$. The top row corresponds to the decoupled case, while the bottom row shows the coupled evolution. In the coupled case, the odd-parity sector is truncated at $\ell_{\max}=5$, while the even-parity sector is truncated at $\ell_{\max}=4$. Left column: $q/q_c=0.5$. Right column: $q/q_c=1.08$. Because of mode coupling, the time-domain signal contains contributions from all multipoles within a given parity sector, up to $\ell_{\max}$. Consequently, the late-time dynamics is dominated by higher-multipole contributions.}
\label{TimeDomain}
\end{figure*}
\begin{figure*}[htb]
\centering
\includegraphics[scale=0.92]{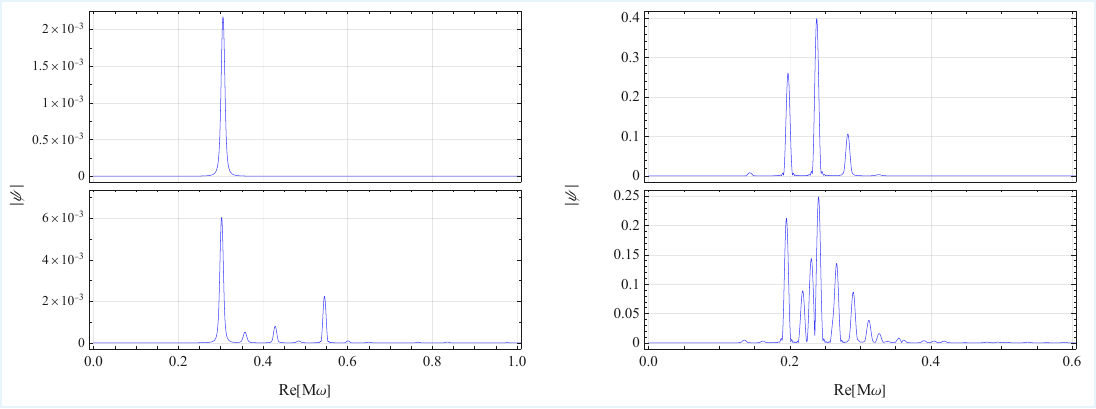}
\caption{Fourier spectrum of the simulation data shown in Fig.~\ref{TimeDomain} for $\ell=1$ and $t/M\in [500,2500]$. Far from the critical charge (left), the dominant decoupled mode (top) is $\omega_{011}=0.3057-0.0042\,\mathrm{i}$. In the coupled case, the spectrum receives contributions from the $\ell=3$ and $\ell=5$ families, leading to additional peaks in the bottom panels. The most relevant frequencies in the observed range are $\Omega^{1}_{01}=0.3035-0.0031\,\mathrm{i}$ and another mode around $0.54-5\times10^{-4}\,\mathrm{i}$. However, the simulation time is not long enough to assign this mode unambiguously to a given $\ell$-family. Close to the critical region (right), $q/q_c\in I_c$, the imaginary parts become so small that the spectrum can be effectively regarded as quasi-degenerate.}.\label{FFT_lmax1_lmax5_fig}
\end{figure*}
In this section, we present the results obtained from a time-domain analysis of the system described by (\ref{eqwave}). We consider the general case where different modes are coupled in the region $r\sim1/B$. In our numerical integration, we consider as an initial condition a Gaussian wave packet, $\psi_\ell(0,r)=\sqrt{2/\pi} \exp{[-\kappa (r-r_0)^2}]\delta_{1\ell}$, falling into the BH. Spatial derivatives are approximated using a fourth-order accurate central finite difference scheme, while time evolution is computed via a second-order explicit method. The corresponding finite difference expressions for the second-order derivatives in time and space are given by
\begin{eqnarray}
\label{eq:Sim:DiscreteDerivativer_*}
\partial_{r_*}^2 \psi_{i,j} &=& \frac{1}{12 \Delta r_*^2} \Bigl( 
-\psi_{i -2, j} + 16 \psi_{i -1, j} - 30 \psi_{i , j} \Bigr.\nonumber \\
& & \quad \quad  \quad \quad  \quad \quad   
+\Bigl. 16 \psi_{i +1, j} - \psi_{i +2, j} \Bigr)
\end{eqnarray}

and
\begin{equation}   \label{eq:Sim:DiscreteDerivativeT}
    \partial_t^2 \psi_{i , j}  = \frac{1}{\Delta t^2} \Bigl(\psi_{i , j+1}  - 2 \psi_{i , j}  + \psi_{i , j-1}  \Bigl).
\end{equation}

As noted in Section~\ref{III}, the perturbation remains confined to a finite spatial region due to the asymptotic behavior of the effective potential for large $Br$. 
This means that very close to the black hole, the perturbation oscillations resemble those of the massive Schwarzschild QNMs, where the effective mass is given by 
$\mu_{\mathrm{eff}}^2 = \mu^2 + 2 |q m B|$. 
Nevertheless, after a time of order $t \sim 1/B$, the wave interacts with the effective wall barrier, producing long-lived modes and exhibiting an echo-like structure, as shown in Fig.~\ref{TimeDomain}. 

Due to the presence of many modes, extracting the complete quasinormal spectrum from the time-domain data is, in general, very challenging. This difficulty becomes even more severe near the critical regime, where the imaginary part of the frequency becomes very small. For this reason, in most cases we extract only the real part of the frequencies through a Fourier analysis of the time-domain signal. In some special cases, however, we employ Prony's method to numerically extract the complete QNM frequencies; see Fig.~\ref{FFT_lmax1_lmax5_fig}.

For the sake of illustration, in Fig.~\ref{FFT_lmax1_lmax5_fig} we show the real parts of the frequencies for $MB=0.1$ at two different charges, one far from and one close to $q_c$, using multipole cutoffs $\ell_{\mathrm{max}}=1$ and $5$. In the coupled case, the spectrum becomes richer, but the fundamental mode, associated with $\ell=1$, already present in the decoupled case remains present, although slightly deformed. Moreover, at late times the coupled signal becomes dominated by higher-multipole modes. This can be seen explicitly in Fig.~\ref{TimeDomain}, where around $t/M \sim 2\times10^3$ the dominant behavior changes from the $\ell=1$ family to a higher-multipole branch with dominant frequency \(\omega\sim0.54-5\times10^{-4} \mathrm{i}\).

Notice also that the same initial Gaussian packet excites more modes as one gets close to the critical charge. This is expected because, in the limit $q\to q_c$, the effective potential changes, leading to further displacement outward of the effective barrier into the region $r\gg 1/B$. This provides an intuitive explanation for the very small decay rates found in this regime. Moreover, we observe that the coupled case yields decay rates much smaller than those of the decoupled case, which strongly suggests the approach to a very long-lived modes, marked by
\begin{equation}
    \lim_{q\to q_c} \mathrm{Im}[\omega] \approx 0.
\end{equation}
Across the transition at $q\approx q_c$, we observe a discontinuity in the spectrum accompanied by the emergence of new modes. For instance, in Fig.~\ref{FFT_lmax1_lmax5_fig} (right), a new fundamental mode appears at \(\mathrm{Re}[M\omega_{011}] \approx 0.14\).

\section{Discussion}\label{IV}
\subsection{Scaling law}
The wave propagating around the Ernst BH is effectively
confined by the asymptotic behavior of the spacetime at infinity.
This setup leads to a dynamics analogous to that of a small perfect
absorber placed inside a confining box. It was shown in
Ref.~\cite{Brito_2014} that, in such systems, the real part of
the QNM frequency scales inversely with the
effective box radius $r_0$. In our analysis, we find that the presence of charge introduces a second characteristic length scale, which displays a critical dependence on the scalar field charge, given by
\begin{equation}
  d\sim \lvert q - q_c \rvert^{-\gamma}.
\end{equation}

To see this, we consider the characteristic length scale \(d\) at which
the quadratic growth
\begin{equation}
  V_{\mathrm{eff}}(r,q_c) \sim B^{4} r^{2}
\end{equation}
starts to compete with the quartic contribution induced by a small
deviation of the charge, $ q = q_c + \epsilon$, with $\qquad \lvert \epsilon \rvert \ll 1$.
In other words, we evaluate
\begin{equation}
  \delta V_{\mathrm{eff}}(d;\epsilon) :=
  \bigl\lvert
    V_{\mathrm{eff}}(d; q_c + \epsilon)
    - V_{\mathrm{eff}}(d; q_c)
  \bigr\rvert
  \sim B^{4} d^{2}.
\end{equation}
The relation above leads to the following scaling law for the
characteristic length ($d$):
\begin{equation}
  d \sim |\epsilon|^{-1/2}
      \left(
        a_0 + a_1 \epsilon + O(\epsilon^{2})
      \right),
\end{equation}
where $a_0$ and $a_1$ are coefficients that depend on the
background parameter $B$. This length diverges as $\epsilon \to 0$, indicating critical behavior with a scaling exponent of 1/2. 
Therefore, we expect the following scaling law for the real frequency,
\begin{equation}
\mathrm{Re}[\omega] \sim |q-q_c|^{1/2}.
\end{equation}
Notice that this analysis is mode-independent. Thus, it suggests universality and is consistent with the scaling we found numerically in Sec.~\ref{IIIb}. The $\sim 8\%$ error is due to the lack of numerical data in the critical region $I_c$.

\subsection{Final Remarks}
Throughout this work, we have argued that charged scalar fields near magnetized BH can undergo a phase transition when they reach a critical charge. Conceptually, this transition is characterized by a change from a confined state, in which the waves remain close to the BH, to a deconfined phase in which the wave escapes to regions far from the BH. Quantitatively, we find strong numerical evidence for critical behavior, marked by a universal critical exponent of approximately $1/2$ for the real frequency. However, although we do not yet have sufficient numerical evidence for universal behavior in the imaginary part, we consistently observe a strong suppression of $|\mathrm{Im}\,\omega|$ as the critical charge is approached, suggesting a genuinely special state of the system. In this way, the results presented in this work provide an interesting example of nontrivial behavior obtained from a purely linear analysis of interaction between magnetized BH and charged field. This also suggests that immediate extensions, such as including BH rotation or backreaction effects may broaden these results and lead to possible real applications, given the frequent occurrence of systems involving black holes and magnetic fields in our Universe.

Any application of the ES solution to astrophysical situations must be treated with caution due to its non-asymptotically flat behavior at infinity. Nevertheless, accretion discs can sustain large scale magnetic fields \cite{Narayan_2003}, thereby magnetizing a black hole. In this spirit, the ES solution may be used at least as an approximation up to some finite distance $L$, especially in scenarios where such magnetic fields provide a confinement/trapping mechanism \cite{romanova1998dynamics}. With these caveats in mind, we can convert our results to physical units, as discussed in this work.

We perform a detailed analysis in the regime $MB\le 10^{-1}$. Restoring SI units, the corresponding magnetic field scale is
\begin{equation}
B_{\rm phys}\sim 10^{14}\left(\frac{M_\odot}{M}\right)\,\mathrm{T},
\end{equation}
where $M_\odot$ is the solar mass. This is far above the largest fields inferred for magnetars, typically
$B_{\rm mag}\sim 10^{-4}B_{\rm phys}$.

One of the most important result of this work is that the critical frequency has a non null value given by Eq.~(\ref{criticalomega}). We interpret this fact as the characteristic frequency of a signal that can propagate to large distances. Therefore, the phase transition can be thought of as an emission threshold limit in which the scalar waves are transmitted. The frequency of this signal for magnetic fields of order $B_{\rm mag}$, has the following magnitude
\begin{equation}
\nu_c \sim 3.2\times10^{-1}\left(\frac{M_\odot}{M}\right)\,\mathrm{Hz}.
\end{equation}
This is negligibly small for stellar mass BH, but it can be larger for small mass. In particular, light primordial black holes are considered as a fraction of dark matter and their interaction with a neutron star is considered as a possible FRB engine \cite{Fuller_2017,abramowicz2018collisions}. For instance, taking $M\sim 10^{23}\,\mathrm{g}$ \cite{Auffinger_2023} yields\footnote{However, it is not clear what the mass of a primordial Black Hole is. For lighter primordial BH the critical frequency can raise this figure to ultraviolet range.} $\nu_c\sim \mathrm{GHz}$. In that case, we come near the size of a progenitor of a Fast Radio Burst \cite{zhang2020physical}. We conclude that such simple models might cope with more sophisticated phenomena \cite{dos2024bingo,Zhang:2024bar}.

\section*{Acknowledgments}
The authors acknowledge financial support from CNPq, Grants No. 303592/2020-6 (E.A.), 141276/2021-5 (M.R.R.), and 141310/2024-3 (E.C.R.). The authors also thank Alberto Saa for carefully reading the manuscript and for helpful discussions. The authors thank the anonymous referee for the careful reading of the manuscript and for constructive comments and suggestions, which significantly improved the present work.


\begin{thebibliography}{}

\end{thebibliography}


\begin{thebibliography}{99}

\bibitem{schwarzschild1999gravitationalfieldmasspoint}
K.~Schwarzschild,
On the gravitational field of a mass point according to Einstein's theory,
\href{https://arxiv.org/abs/physics/9905030}{arXiv:physics/9905030}.

\bibitem{stephani2009exact}
H.~Stephani, D.~Kramer, M.~MacCallum, C.~Hoenselaers, and E.~Herlt,
\textit{Exact Solutions of Einstein's Field Equations}
(Cambridge University Press, Cambridge, 2009).

\bibitem{griffiths2009exact}
J.~B.~Griffiths and J.~Podolsk\'y,
\textit{Exact Space-Times in Einstein's General Relativity}
(Cambridge University Press, Cambridge, 2009).

\bibitem{Abbott_2016}
B.~P.~Abbott \textit{et al.},
Phys.\ Rev.\ Lett.\ \textbf{116}, 061102 (2016),
\href{https://arxiv.org/abs/1602.03837}{arXiv:1602.03837}.

\bibitem{gravitationalwavesfrominflation_2016}
M.~C.~Guzzetti, N.~Bartolo, M.~Liguori \textit{et al.},
Riv.\ Nuovo Cim.\ \textbf{39}, 399 (2016),
\href{https://arxiv.org/abs/1605.01615}{arXiv:1605.01615}.

\bibitem{akiyama2022first}
Event Horizon Telescope Collaboration, K.~Akiyama, A.~Alberdi \textit{et al.},
Astrophys.\ J.\ Lett.\ \textbf{930}, L14 (2022),
\href{https://arxiv.org/abs/2311.09479}{arXiv:2311.09479}.

\bibitem{hallinan2017radio}
G.~Hallinan, A.~Corsi, K.~Mooley \textit{et al.},
Science \textbf{358}, 1579 (2017),
\href{https://arxiv.org/abs/1710.05435}{arXiv:1710.05435}.

\bibitem{abdalla2022bingo}
E.~Abdalla, E.~G.~Ferreira, R.~G.~Landim \textit{et al.},
Astron.\ Astrophys.\ \textbf{664}, A14 (2022),
\href{https://arxiv.org/abs/2107.01633}{arXiv:2107.01633}.

\bibitem{dos2024bingo}
M.~V.~dos Santos, R.~G.~Landim, G.~A.~Hoerning \textit{et al.},
Astron.\ Astrophys.\ \textbf{681}, A120 (2024),
\href{https://arxiv.org/abs/2308.06805}{arXiv:2308.06805}.

\bibitem{blandford1982hydromagnetic}
R.~D.~Blandford and D.~G.~Payne,
Mon.\ Not.\ R.\ Astron.\ Soc.\ \textbf{199}, 883 (1982).

\bibitem{zhang2020physical}
B.~Zhang,
Nature \textbf{587}, 45 (2020),
\href{https://arxiv.org/abs/2011.03500}{arXiv:2011.03500}.

\bibitem{Bochenek_2020}
C.~D.~Bochenek, V.~Ravi, K.~V.~Belov \textit{et al.},
Nature \textbf{587}, 59 (2020),
\href{https://arxiv.org/abs/2005.10828}{arXiv:2005.10828}.

\bibitem{korobkin2011stability}
O.~Korobkin, E.~B.~Abdikamalov, E.~Schnetter \textit{et al.},
Phys.\ Rev.\ D \textbf{83}, 043007 (2011),
\href{https://arxiv.org/abs/1011.3010}{arXiv:1011.3010}.

\bibitem{RevModPhys.70.1}
S.~A.~Balbus and J.~F.~Hawley,
Rev.\ Mod.\ Phys.\ \textbf{70}, 1 (1998).

\bibitem{konoplya2006stability}
R.~A.~Konoplya and A.~Zhidenko,
Phys.\ Rev.\ D \textbf{73}, 124040 (2006),
\href{https://arxiv.org/abs/gr-qc/0605013}{arXiv:gr-qc/0605013}.

\bibitem{Brito_2014}
R.~Brito, V.~Cardoso, and P.~Pani,
Phys.\ Rev.\ D \textbf{89} (2014),
\href{https://arxiv.org/abs/1405.2098}{arXiv:1405.2098}.

\bibitem{Filho:2024ilq}
A.~A.~A.~Filho, K.~Jusufi, B.~Cuadros-Melgar \textit{et al.},
Phys.\ Dark Universe \textbf{46}, 101711 (2024),
\href{https://arxiv.org/abs/2401.15211}{arXiv:2401.15211}.

\bibitem{Lin:2019fte}
K.~Lin, Y.~Liu, W.~L.~Qian \textit{et al.},
Phys.\ Rev.\ D \textbf{100}, 065018 (2019),
\href{https://arxiv.org/abs/1909.04347}{arXiv:1909.04347}.

\bibitem{Zhu_2014}
Z.~Zhu, S.~J.~Zhang, C.~Pellicer \textit{et al.},
Phys.\ Rev.\ D \textbf{90} (2014),
\href{https://arxiv.org/abs/1405.4931}{arXiv:1405.4931}.

\bibitem{Ribeiro:2024jkm}
E.~C.~Ribeiro, L.~Formigari, M.~R.~Ribeiro, Jr.\ \textit{et al.},
Phys.\ Rev.\ D \textbf{111}, 024043 (2025),
\href{https://arxiv.org/abs/2411.11117}{arXiv:2411.11117}.

\bibitem{de_Freitas_2017}
V.~P.~de Freitas and A.~Saa,
Phys.\ Rev.\ D \textbf{95} (2017),
\href{https://arxiv.org/abs/1703.10883}{arXiv:1703.10883}.

\bibitem{aly2025more}
F.~Aly, M.~A.~Mansour, and D.~Stojkovic,
Phys.\ Rev.\ D \textbf{111}, 104082 (2025),
\href{https://arxiv.org/abs/2410.12775}{arXiv:2410.12775}.

\bibitem{1976JMP....17...54E}
F.~J.~Ernst,
J.\ Math.\ Phys.\ \textbf{17}, 54 (1976).

\bibitem{shaymatov2022constraints}
S.~Shaymatov, M.~Jamil, K.~Jusufi \textit{et al.},
Eur.\ Phys.\ J.\ C \textbf{82}, 636 (2022),
\href{https://arxiv.org/abs/2205.00270}{arXiv:2205.00270}.

\bibitem{nayak1997gyroscopic}
K.~R.~Nayak and C.~Vishveshwara,
Gen.\ Relativ.\ Gravit.\ \textbf{29}, 291 (1997).

\bibitem{konoplya2008quasinormal}
R.~A.~Konoplya and R.~Fontana,
Phys.\ Lett.\ B \textbf{659}, 375 (2008),
\href{https://arxiv.org/abs/0707.1156}{arXiv:0707.1156}.

\bibitem{Horowitz_1997}
G.~T.~Horowitz and H.~J.~Sheinblatt,
Phys.\ Rev.\ D \textbf{55}, 650 (1997),
\href{https://arxiv.org/abs/gr-qc/9607027}{arXiv:gr-qc/9607027}.

\bibitem{choptuik1993universality}
M.~W.~Choptuik,
Phys.\ Rev.\ Lett.\ \textbf{70}, 9 (1993).

\bibitem{werneck2021nrpycritcol}
L.~R.~Werneck, Z.~B.~Etienne, E.~Abdalla \textit{et al.},
Class.\ Quantum Grav.\ \textbf{38}, 245005 (2021),
\href{https://arxiv.org/abs/2106.06553}{arXiv:2106.06553}.

\bibitem{gralla2018scaling}
S.~E.~Gralla and P.~Zimmerman,
J.\ High Energy Phys.\ \textbf{06} (2018) 061,
\href{https://arxiv.org/abs/1804.04753}{arXiv:1804.04753}.

\bibitem{becar2023quasinormal}
R.~B\'ecar, P.~Gonz\'alez, and Y.~V\'asquez,
Eur.\ Phys.\ J.\ C \textbf{83}, 75 (2023),
\href{https://arxiv.org/abs/2211.02931}{arXiv:2211.02931}.

\bibitem{Melvin:1963qx}
M.~A.~Melvin,
Phys.\ Lett.\ \textbf{8}, 65 (1964).

\bibitem{thorne1965absolute}
K.~S.~Thorne,
Phys.\ Rev.\ \textbf{139}, B244 (1965).

\bibitem{wild1980surface}
W.~J.~Wild and R.~M.~Kerns,
Phys.\ Rev.\ D \textbf{21}, 332 (1980).

\bibitem{dadhich1979trajectories}
N.~Dadhich, C.~Hoenselaers, and C.~Vishveshwara,
J.\ Phys.\ A: Math.\ Gen.\ \textbf{12}, 215 (1979).

\bibitem{stuchlik1999photon}
Z.~Stuchl\'ik and S.~Hled\'ik,
Class.\ Quantum Grav.\ \textbf{16}, 1377 (1999),
\href{https://arxiv.org/abs/0803.2536}{arXiv:0803.2536}.

\bibitem{pani2013advanced}
P.~Pani,
Int.\ J.\ Mod.\ Phys.\ A \textbf{28}, 1340018 (2013),
\href{https://arxiv.org/abs/1305.6759}{arXiv:1305.6759}.

\bibitem{Narayan_2003}
R.~Narayan, I.~V.~Igumenshchev, and M.~A.~Abramowicz,
Publ.\ Astron.\ Soc.\ Jpn.\ \textbf{55}, L69 (2003),
\href{https://arxiv.org/abs/astro-ph/0305029}{arXiv:astro-ph/0305029}.

\bibitem{romanova1998dynamics}
M.~Romanova, G.~Ustyugova, A.~Koldoba \textit{et al.},
Astrophys.\ J.\ \textbf{500}, 703 (1998).

\bibitem{Fuller_2017}
G.~M.~Fuller, A.~Kusenko, and V.~Takhistov,
Phys.\ Rev.\ Lett.\ \textbf{119}, 061101 (2017),
\href{https://arxiv.org/abs/1704.01129}{arXiv:1704.01129}.

\bibitem{abramowicz2018collisions}
M.~A.~Abramowicz, M.~Bejger, and M.~Wielgus,
Astrophys.\ J.\ \textbf{868}, 17 (2018),
\href{https://arxiv.org/abs/1704.05931}{arXiv:1704.05931}.

\bibitem{Auffinger_2023}
J.~Auffinger,
Prog.\ Part.\ Nucl.\ Phys.\ \textbf{131}, 104040 (2023),
\href{https://arxiv.org/abs/2206.02672}{arXiv:2206.02672}.

\bibitem{Zhang:2024bar}
X.~Zhang, Y.~Sang, G.~A.~Hoerning \textit{et al.},
Astrophys.\ J.\ \textbf{991}, 189 (2025),
\href{https://arxiv.org/abs/2411.17516}{arXiv:2411.17516}.

\end{thebibliography}
\end{document}